# HIGH DENSITY $MgB_2$ OBTAINED BY REACTIVE LIQUID Mg INFILTRATION

GIOVANNI GIUNCHI

*EDISON S.p.a. – Divisione Ricercs e Sviluppo – Via U. Bassi 2 20159 Milano (Italy)*
*E-mail: giovanni.giunchi@edison.it*

A new route to the sintering of $MgB_2$ has been identified, based on a reactive infiltration of liquid Mg on a powdered B preform. The technique allows to obtain large bulk manufacts in an inexpensive way, without the need of high pressure apparatus. The best of the obtained samples shows a transport current density of 3 $kA/cm^2$ at 4.2K and 9 T. The critical aspects of the technology are presented , together with the recent achievements and the perspective applications.

## 1 Introduction

After the discovery of the superconductivity of the $MgB_2$[1], about one year ago, many works have been devoted to improve the superconductive characteristics of this material in its various shapes, either bulk or tape or thin film. Focusing on the bulk material, the most efficient sintering technique used to densify the products was the hot pressing, either of the $MgB_2$ powders or of the mixture of the Magnesium and Boron elemental precursors. In both cases the high pressure prevents the decomposition of the $MgB_2$, that at normal pressure occurs already at temperature lower than 1000°C [2]. The high pressure sintering, from the industrial point of view, can be realized either as uniaxial hot pressing(UHP) or as hot isostatic pressing (HIP). Nevertheless both techniques, even if feasible, show some limitations in the dimension of the manufacts , due to the practical difficulty in the use of very large vessels and dyes, when high pressures and high temperatures are involved.
Looking for alternative $MgB_2$ synthesis routes, we have found that the reaction of B and Mg, to produce $MgB_2$ can be efficiently performed, if conducted in an appropriate way[3], producing objects very dense and with very high superconducting characteristics, without the need of high pressure apparatus.
In the following, the main features of the new sintering method are described together with the microscopic characterization of the resulting products and with the evaluation of their transport current capability in high magnetic fields. The high values of the critical current density are a signature of the quality of the material and of the possibility of its use in future applications.
In this respect we review briefly some of the most appealing superconducting power applications, which can benefit of the use of the $MgB_2$ in bulk form and of the use of refrigeration apparatus in a medium temperature range ( up to about 30K), typical of the superconductive properties of the $MgB_2$ material.

## 2 Reactive liquid Mg infiltration

The thermodynamics of the phases in the Mg – B system has been fully described, with the help of the CALPHAD computing program, by Liu et al.[4]. A relevant diagram for the $MgB_2$ reactive sintering, according to the present findings, is given in Figure 1.

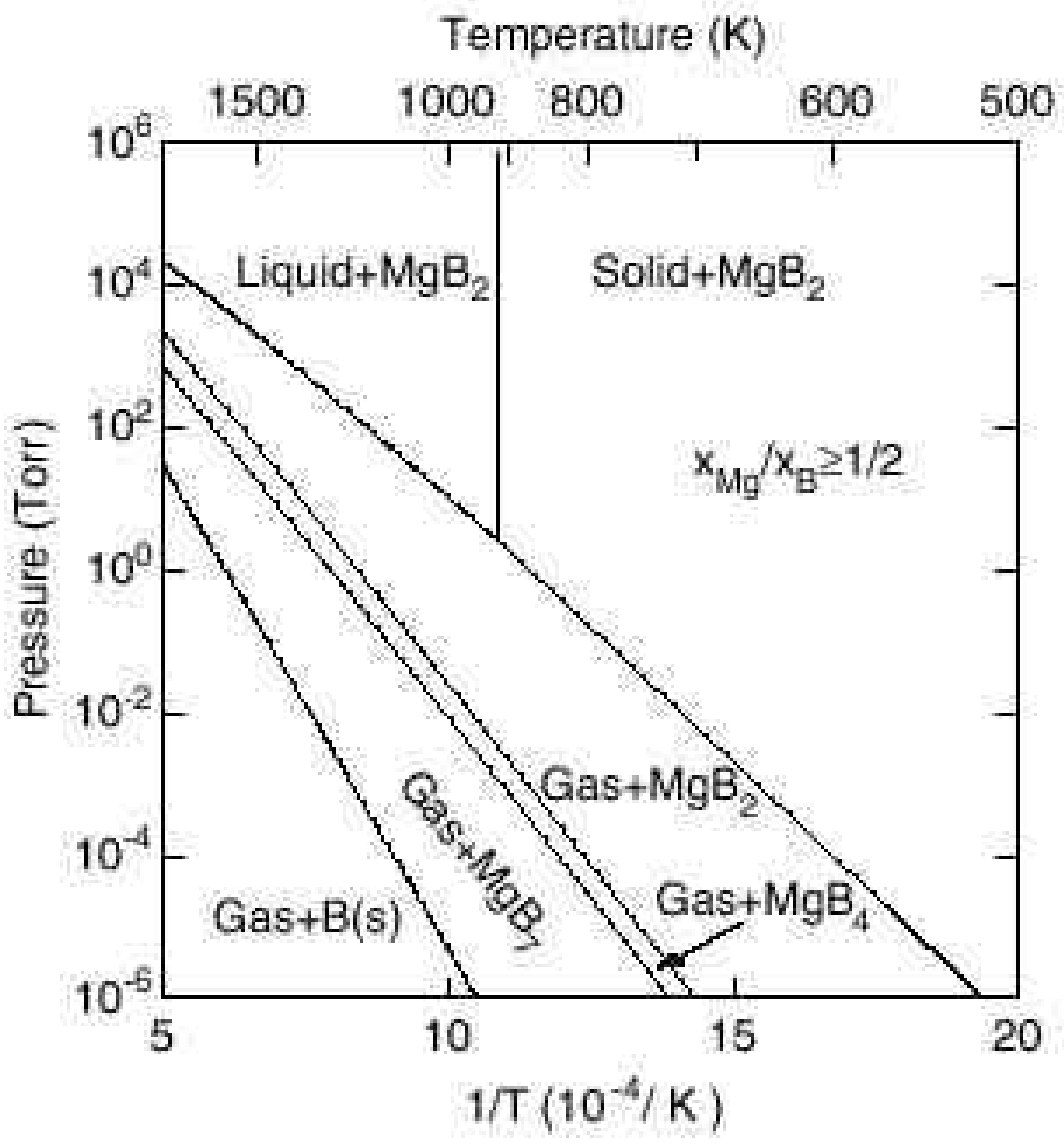

**Figure 1 :** Computed phase diagram for Mg-B system , with molar Mg> 0.5 B(After Liu et al.[4])

In this diagram the region of co-existence of the $MgB_2$ with a liquid phase is the place where we have verified that the reaction occurs in a more efficient way.
We also verified that the $MgB_2$ formation and its densification occur simultaneously, when a porous body build up of B powders (preform) is infiltrated with the liquid magnesium at the appropriate pressures and temperatures.
Other general conditions appropriated for the formation of high density $MgB_2$ are:

1. a Boron preform made by crystalline powders of medium–high grain size (mean value between 30 and 70 μm) , having a green density of at least 50% of the full density of B.
2. an amount of Mg larger than the stoichiometry for the $MgB_2$ formation, added in a massive form, i.e. not mixed with the B powders.
3. a sealed metallic container, lined with materials not or slightly reacting with Mg or B (for example steel or niobium), in which the B preform is tightly inserted together with the massive body of Mg, in such a way that the inner walls of the container counteract the pressure produced by the expansion of the $MgB_2$ forming phase.

With respect to the standard way of mixing B and powders or chips of Mg and of sintering both in an high pressure apparatus like UHP or HIP, the presented technology is able to guarantee densification of the products without external pressure on the container.
As counterproof, when we have applied our conditions to a mixture of Mg and B powders, instead of the combination of the B preform with the Mg massive body, the resulting densification of the reaction products was very poor, due to the inability of the system to recover the microvoids left by the Mg.
Surprisingly, when the appropriate B powders are packed in a preform with the right green density, the liquid Mg is able to percolate the preform, also against the gravity field. So, even if we put a Mg reservoir on the bottom of the container, just below the preform, we find, after reaction, that the Mg has been absorbed by the B and has reacted completely up to the stoichiometric amount for the $MgB_2$ formation.
One of the characteristics of our best densification technology is the generation, in the final manufacts, of large voids corresponding to the zones where the Mg bodies were inserted before of the reaction. The presence of these voids, together with their location, can be utilized for the applicative needs in the designing of the superconductive manufacts.
A further advantage of the infiltration technology is that the B powders remain stationary, during the reaction, allowing to obtain objects of predetermined shape that require only a surface finishing.

## 3 Manufacts fabrication

The manufacturing technology should include the following steps:

- fill a metallic container with the B powders and with an appropriate amount of Mg (either as solid body or via liquid infiltration)
- weld the container with an high temperature resistant technique , like the Tungsten Inert Gas (TIG)
- anneal the container in a conventional furnace
- extract the reacted $MgB_2$ object from the container
- finish the surface of the object and cut to the desired shape

As examples of this manufacturing procedure, two typical objects are presented:
A)a cylindrical pellet and B) an hollow cylinder .

Both manufacts are made using 99.9% pure Mg ingots and a crystalline Boron powder (H.S. Starck) of 99.5% of purity, grinded and sieved with a 100 μm sieve .
The pellet (A) has been obtained by inserting the Mg and the B powders in a steel container ($\phi_{int}$=17 mm), with internal lining of Nb foil.

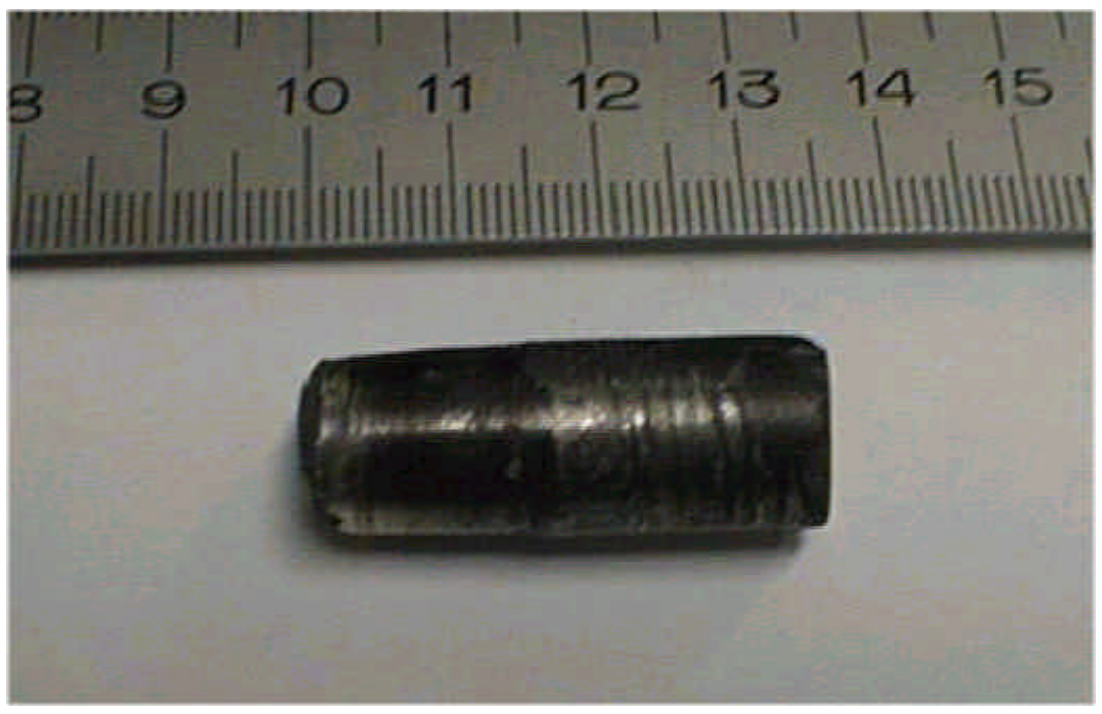

**Figure 2** – Cylindrique pellet (A) of $MgB_2$

The thickness of the wall of the container was 4 mm. Mg was added in form of two solid cylinders, located one above and one below the B powders, corresponding to a total molar ratio Mg/B=0.633.
After welding of the steel container in an Ar atmosphere, the same was annealed at 950°C , for 3 hours, in vertical position and then naturally cooled up to R.T.

The $MgB_2$ pellet(A), extracted from the steel container, shown in Figure 2, has a density of 2.40 g/cm$^3$ .
The second manufact, the hollow cylinder (B), has been obtained inserting the B powders between the inner wall of a steel cylinder, lined with Nb foil, and a coaxial internal Mg solid cylinder .

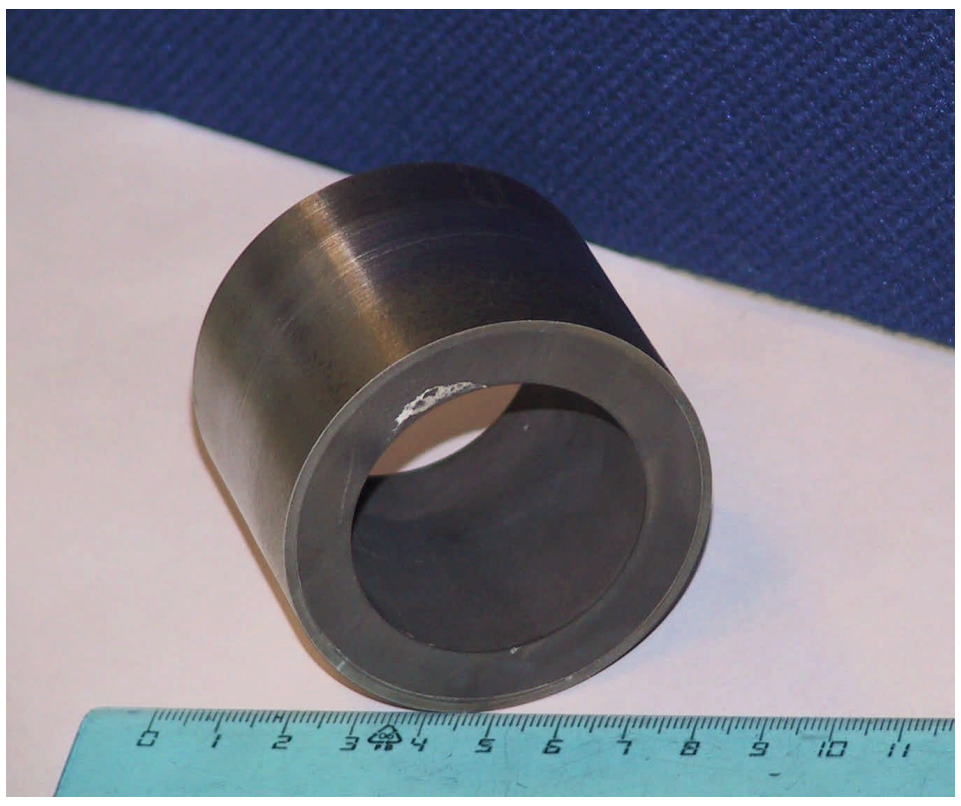

**Figure 3** : $MgB_2$ hollow cylinder (B), formed inside a steel container

The molar ratio Mg/B was = 0.553. After the welding, the steel container was annealed in a vertical position at 850°C , for 3 hours, than cooled naturally up to R.T.
The internal finishing of the obtained $MgB_2$ hollow cylinder, shown in Figure 3, was performed by removing, by electro-erosion, a surface layer of about 1 mm in thickness, made up by irregular material.

## 4 Microscopic morphology and superconductive properties

### *4.1 Morphology*

The bulk $MgB_2$ compacts have been observed by optical (OM) and scanning electron microscopy (SEM) and the relevant feature of their microstructure is the presence of large grains embedded between smaller ones, as reported in Figure 4. The larger grains are reminiscent of the dimensions of the starting Boron crystalline powders and their structure, even if very dense , is characterized by a finer granularity and by many fractures ,as displayed in Figure 5.

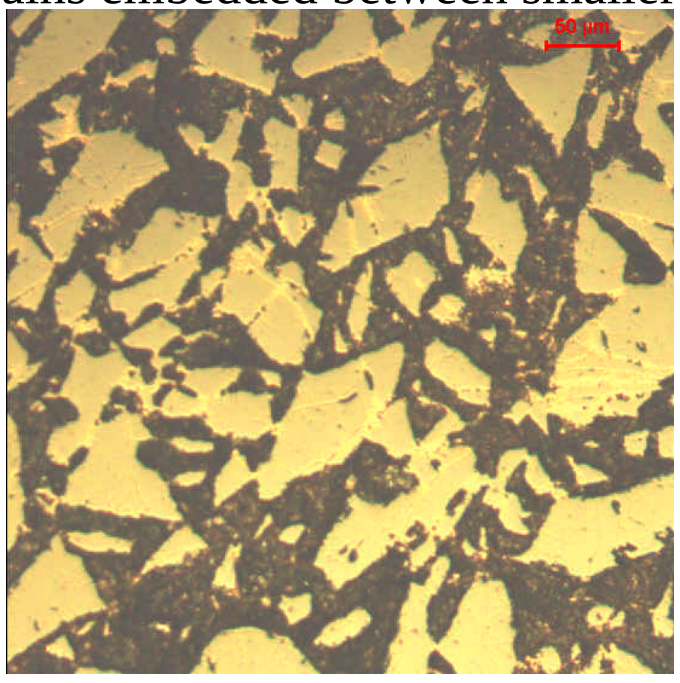

**Figure 4 :** Optical micrograph of a polished surface of sample (A)

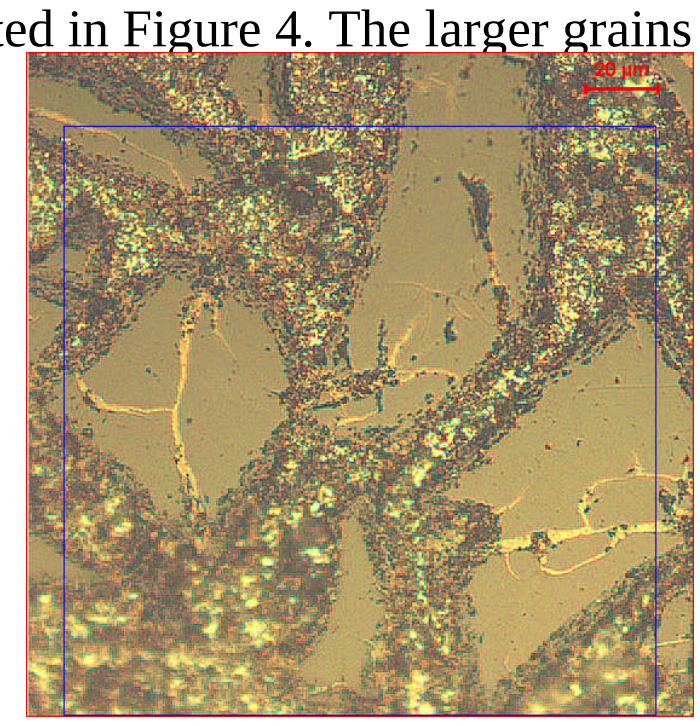

**Figure 5 :** Optical micrograph showing intragrain fractures in sample (A)

Even if the fractures and the grain boundaries don't stop the superconductive currents, the volume proportion of the larger grains with respect to the smaller ones is one of the parameters which determine the current transport performances of the superconductor [5].

After x-ray fluorescence microanalysis on the sample a larger presence of metallic Mg is detected both in the region of the small grains and inside the fractures of the larger grains. The overall presence of the metallic Mg in the bulk has been also detected by x-ray powder diffraction analysis, as reported in the diagram of Figure 6.

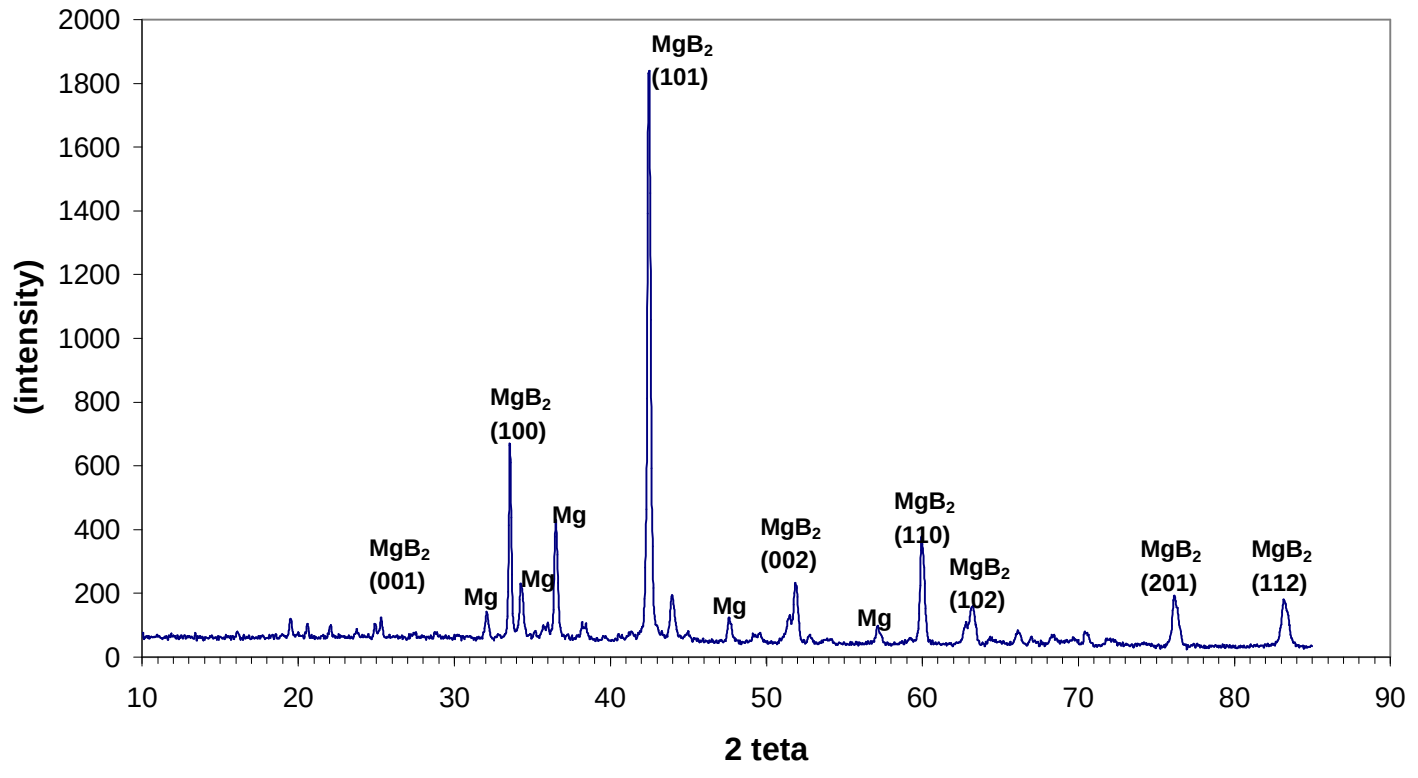

**Figure 6** : X-ray powder diffraction pattern of the sample (A)

## *4.2 Superconductive properties*

The most astonished property of the manufactured $MgB_2$ compacts is the relative insensitivity of their granularity on the superconductive current percolation. A first observation of this characteristics, on our samples, was reported by Gozzelino et al. [6] with magneto optical microscopy . This analysis showed that the magnetic field, applied to low temperature, was completely repelled from the inner part of the manufact: a behaviour that is typical of single crystals. Furthermore on one sample densified as above it was found the first experimental evidence of the Josephson effect for a break junction[7].
The resistively measured Tc was in the range of 39-40 K[8], similar to the best samples up to now published, and the features of the resistivity curve , displayed in Figure 7 , shows a very limited spread at the Tc ( $\Delta T = 0.5$ K) and an RRR value of about 14.

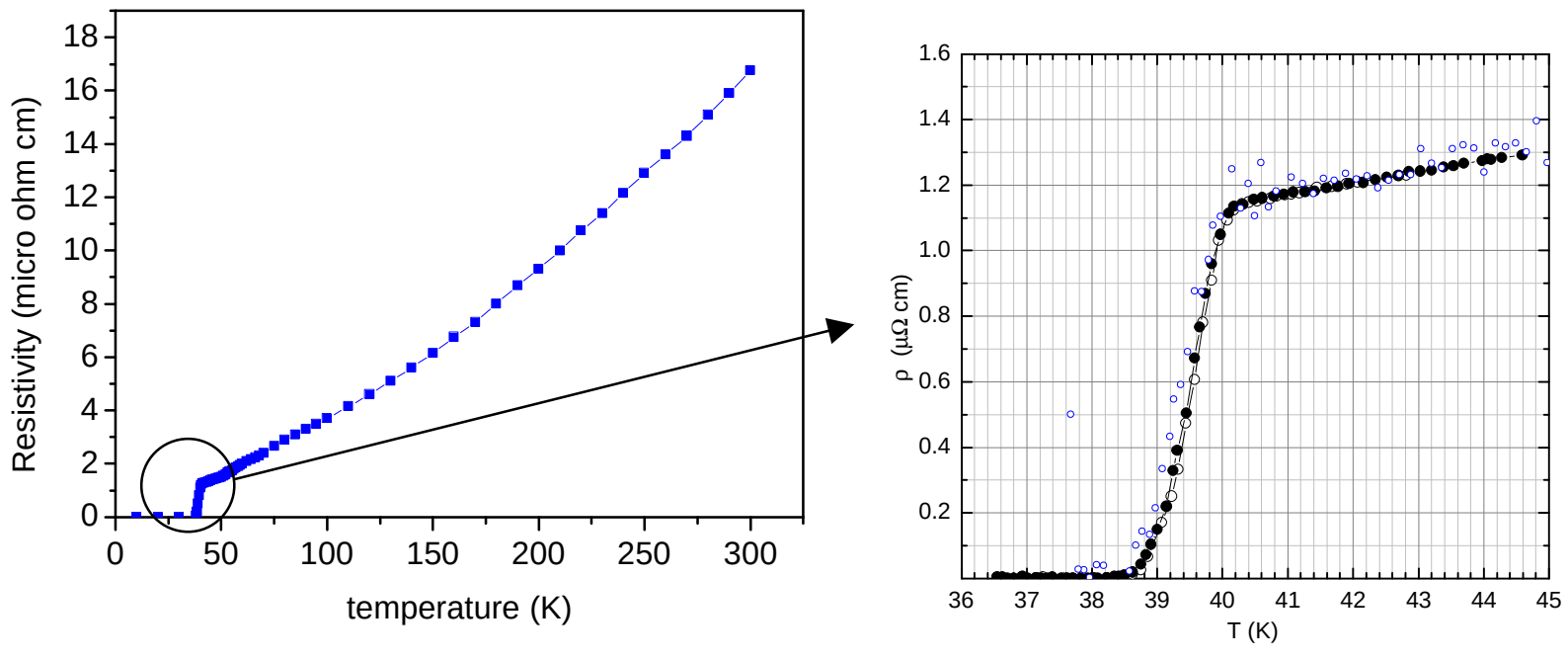

**Figure 7 :** Resistivity versus temperature for sample (A)

For a bar extracted from sample (A) the transport critical current density has been measured at 4.2 K , but only at high magnetic fields, from 9 T up to 12 T, to avoid the thermal degradation of the contacts of the superconducting bar with the current leads. The obtained values of the critical current density are displayed in Figure 8 together with the values of other typical bulk $MgB_2$ superconductors.

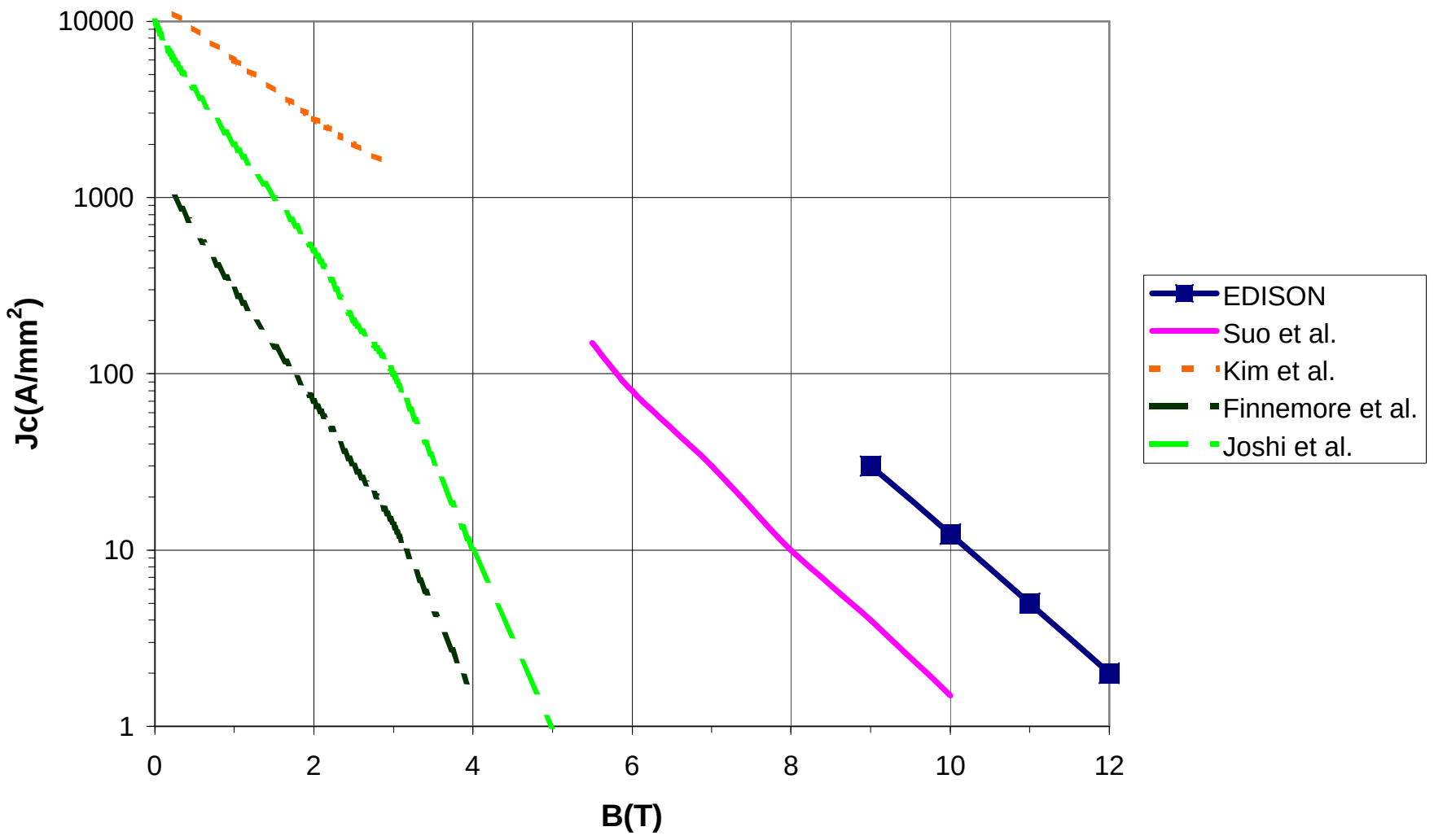

**Figure 8** : Transport critical current densities of the sample (A),measured at 4.2 K, as compared to published values for bulk or thick samples : Suo et al [9], Kim et al [10], Finnemore et al.[11], Joshi et al. [12]

Even if our critical current densities are very high with respect to other bulk $MgB_2$, a comparison with other superconducting materials, as displayed in Figure 9, indicates that the actual samples need the introduction of better flux pinning mechanism to compete in the applications of high magnetic fields ( larger than 4 T), at least at the liquid He temperature.
The temperature behaviour of the critical current density of the presented $MgB_2$ bulk materials has been measured magnetically and reported elsewhere[5]. Even if the comparison between magnetic and transport measurements is questionable an useful indication can be extract as relative values. For example at 2 T, the critical current density at 4.6 K is about 5 time larger than at 20 K , so, linearly extrapolating the data of Figure 8, we can expect that at this temperature, typical for refrigeration with cryocoolers, the critical current density for the bulk should be in the range of several thousands of $A/mm^2$.

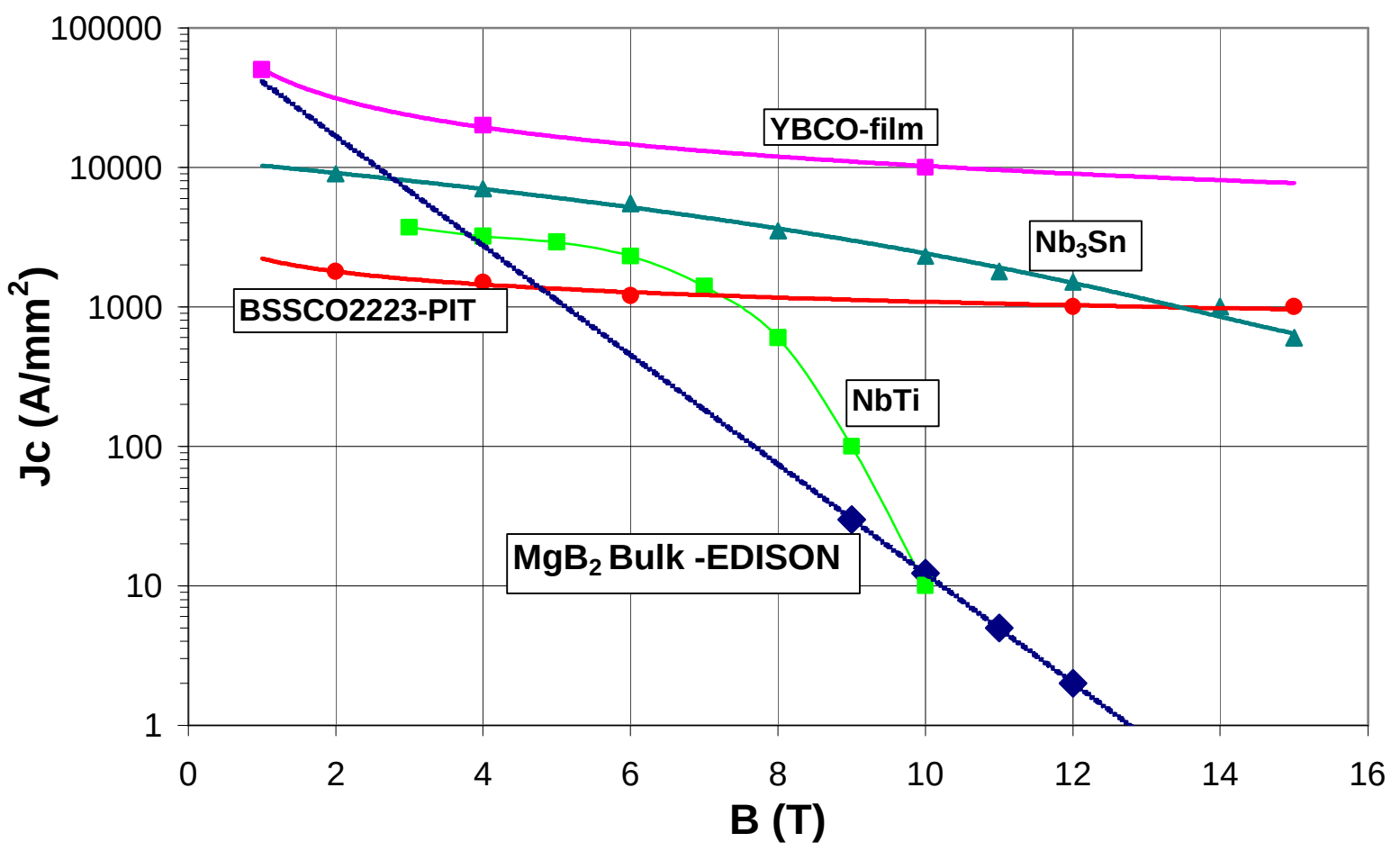

**Figure 9 :** Transport critical current density for bulk $MgB_2$, measured at 4.2 K, compared for other superconducting materials

## 5 Perspective applications of the bulk $MgB_2$ material

The $MgB_2$ bulk material presents several major advantages both with respect to the low temperature superconductors and with respect to the higher temperature superconductors based on cuprates. With respect to NbTi and $Nb_3Sn$ the use of a refrigeration system which can avoid the liquid He is the major advantage for $MgB_2$ . With respect to the cuprates, the $MgB_2$ presents a perspective lower manufacturing cost, a substantially better mechanical behaviour and an intrinsic lower density .

Regarding the refrigeration issues, the coolant capacity of the most interesting liquid refrigerants in the 4 ÷ 77K range, i.e. He,Ne,$N_2$, is substantially in favour of Ne and $N_2$. Indeed the evaporation latent heat for unit volume, of these refrigerants is respectively 2.54, 104.5 and 161 ($J/cm^3$). That means that the Ne is penalized, with respect to $N_2$, by only a factor of 1.5 , a penalty that can be considered minimal in comparison with the other costs of the overall superconducting system.

The power applications of the bulk material should be related to the two main physical characteristics of these superconducting manufacts, i.e. their magnetic trapping ability and their properties to present a variable inductance . As bulk magnets, the main area of applications concern the passive levitated bearings [13], the shielding of alternate and low magnetic fields[14}, the electrical motor either linear or rotating[15]. As conductors or variable resistors or inductors, the main application concern the current lead of cryogenic apparatus and the fault current limiter for the electrical grids, either inductive[16] or resistive[17].

All these applications have been deeply investigated with theoretical models in the case of bulk cuprates, verifying the feasibility of many systems. However, up to now, only few demonstrators of limited power have been built, mainly due to the not enough mechanical strength of the ceramic cuprates and partly due to their limitation in the magnetic performances. In the case of the $MgB_2$ we have indications that these main hurdles for the application of bulk superconducting materials can be overcame , even more motivated by the possibility, here reported, to have a reinforced stainless steel container in several manufacts. The major question mark , at this moment, regards the hypothesis that for such power applications a refrigeration system can be properly designed and economically built.

# 6 Acknoledgement

EDISON acknowledges the Lecco Laboratory of the Milan Section of the CNR-IENI Institute for the availabilty of its manufacturing facilities.